# The Light Curve of Supernova 1987A: The Structure of the Presupernova and Radioactive Nickel Mixing

## V. P. Utrobin[*]

*Institute for Theoretical and Experimental Physics,*
*ul. Bol'shaya Cheremushkinskaya 25, Moscow, 117259 Russia*
*Max-Planck-Institut für Astrophysik, Karl-Schwarzschild-Str. 1, Garching, D-85741 Germany*

**Abstract**—We have studied the influence of the presupernova structure and the degree of $^{56}$Ni mixing on the bolometric light curve of SN 1987A in terms of radiation hydrodynamics in the one-group approximation by abandoning LTE and by taking into account nonthermal ionization and the contribution of spectral lines to its opacity. Our study shows that moderate $^{56}$Ni mixing at velocities $\leq 2500$ km s$^{-1}$ can explain the observed light curve if the density of the outer layers of the presupernova exceeds the value obtained in the evolutionary model of a single nonrotating star severalfold. Abandoning LTE and allowing for nonthermal ionization when solving the equation of state and calculating the mean opacities and the thermal emission coefficient leads to a significant difference between the gas temperature and the radiation temperature in the optically thin layers of the supernova envelope. We demonstrate the fundamental role of the contribution of spectral lines to the opacity in an expanding envelope and of the accurate description of radiative transfer in reproducing the observed shape of the bolometric light curve. We have found that disregarding the contribution of spectral lines to the opacity introduces an error of $\sim 20\%$ into the explosion energy, and that a similar error is possible when determining the mass of the ejected matter. The resonant scattering of radiation in numerous lines accelerates the outer layers to velocities of $\approx 36\,000$ km s$^{-1}$; this additional acceleration affects the outer layers with a mass of $\approx 10^{-6} M_\odot$. Proper calculations of the supernova luminosity require that not only the delay effects, but also the limb-darkening effects be taken into account.

## INTRODUCTION

Supernova (SN) 1987A, which exploded in the Large Magellanic Cloud (LMC), still provides extensive astrophysical information. Nevertheless, it holds many secrets, offering a unique opportunity to study in depth this grandiose event and the preceding evolution of the exploded star. One of the surprises was the fact that the exploded star (presupernova) turned out to be a blue supergiant rather than a red supergiant, as expected for type-II supernovae with a plateau (SN IIP). The explosion of the blue supergiant confirmed the far-sighted conclusion reached by Shklovsky (1984) several years earlier that, since the irregular galaxies (to which the LMC belongs) are metal-poor, the formation of extended stellar envelopes is hampered in presupernovae; therefore, supernovae with properties similar to those shown by SN 1987A must explode in these galaxies instead of ordinary SN II. Indeed, the underabundance of heavy elements in the LMC matter compared to the cosmic composition favors the formation of blue supergiants (Arnett 1987; Hillebrandt *et al.* 1987). In the case of SN 1987A, however, it cannot explain the high nitrogen abundance that was revealed in the circumstellar matter by an analysis of ultraviolet lines (Cassatella 1987; Lundqvist and Fransson 1996). In addition to standard assumptions, either a modification of convective mixing through the meridional circulation induced by rotation of the star during its evolution (Weiss *et al.* 1988), or limited semiconvection at a low abundance of heavy elements (Woosley *et al.* 1988), or invoking evolutionary effects in a close binary system (Podsiadlowski and Joss 1989; Hillebrandt and Meyer 1989) is required to interpret these two facts in evolutionary calculations.

All of these possibilities were equally promising until the ESO ground-based New Technology Telescope (Wampler *et al.* 1990) and the NASA/ESA Hubble Space Telescope (Jakobsen *et al.* 1991) discovered intricate ring structures around SN 1987A. The existence of these structures imposes severe constraints on the pattern of evolution of the presuperno-va, necessitating deviations from spherical symmetry, at least shortly before the supernova explosion. For a single star, this suggests at least including rotation effects. However, including these effects and using new OPAL opacities has not freed the evolutionary calculations from problems in explaining the observed properties of the presupernova, requiring a more thorough development of the physics of rotating stars and convective mixing (Woosley *et al.* 1997). In contrast, the evolution of stars in a close binary is so rich in possibilities that it admits not only the model of accretion of a substantial amount of matter from the secondary component onto the presupernova, but also the model of a complete merger with it (Podsiadlowski 1992).

[*]E-mail: utrobin@itep.ru

Since the structure of the presupernova and the chemical composition of its outer layers (which are unaffected by explosive nucleosynthesis, being the end result of the entire evolution of the star) in many respects determine the pattern of supernovae explosion, the incompleteness of the picture outlined above and the absence of decisive arguments for a particular evolutionary scenario of the presupernova prompt us to analyze the light curve of SN 1987A once again. Interest is also stirred by the fact that the light curve is shaped mainly by radioactive decays of $^{56}$Ni and $^{56}$Co whose distribution in the supernovae envelope is a clear trace left by the explosion mechanism. The hydrodynamic models of SN 1987A based on evolutionary calculations of the presupernova necessitated strong $^{56}$Ni mixing up to velocities of $\sim$4000 km s$^{-1}$ to reproduce the observed bolometric light curve (Woosley 1988; Shigeyama and Nomoto 1990; Blinnikov *et al.* 2000). On the other hand, the hydrodynamic modeling of the SN 1987A explosion that uses nonevolutionary presupernova models and that proceeds from the best agreement with observations has shown the possibility of moderate $^{56}$Ni mixing up to velocities of $\sim$2000 km s$^{-1}$ (Utrobin 1993). Note that Blinnikov *et al.* (2000) not only solved the system of radiation hydrodynamics equations in the multigroup approximation, but also took into account the contribution of spectral lines to the opacity, while other researchers restricted their analyses to the simple approximation of radiative heat conduction.

Our goal here is to investigate the influence of the presupernova structure and the degree of $^{56}$Ni mixing on the bolometric light curve of SN 1987A in terms of radiation hydrodynamics in the one-group approximation by abandoning local thermodynamic equilibrium (LTE) and by taking into account nonthermal ionization and the contribution of spectral lines to the opacity. Our results have confirmed the previous conclusions listed above, provided that the density of the outer layers of the nonevolutionary presupernova model is several times higher than the density in the evolutionary model of a single nonrotating star. We give the system of radiation hydrodynamics equations in the one-group approximation with the equation of state for an ideal gas in a nonequilibrium radiation field and for nonthermal ionization with appropriate mean opacities and thermal emission coefficient. We describe the numerical method for solving this system of equations and consider the hydrodynamic models studied. We compare the evolutionary and nonevolutionary presupernova models, analyze the behavior of the gas temperature and the radiation temperature in the supernova envelope, and study the role of nonthermal ionization, the contribution of spectral lines to the opacity, limb-darkening effects, and the chemical composition of the surface layers in the supernova explosion. In conclusion, we discuss our results and their possible implications.

## RADIATION HYDRODYNAMICS

The CRAB software package has been developed for hydrodynamic studies of supernovae. This package models an unsteady one-dimensional spherically symmetric gas flow in the fields of gravity and nonstationary nonequilibrium radiation in Lagrangian variables. The time-dependent radiative transfer equation, written in a comoving frame of reference to within terms on the order of the ratio of the matter velocity to the speed of light, can be solved as a system of equations for the zeroth and first moments of the radiation intensity in angular variable. To close this system of moment equations, we use a variable Eddington factor that can be calculated by directly taking into account the scattering of radiation in the medium. The satisfaction of the LTE conditions when solving the equation of state and determining the mean opacities and the thermal emission coefficient is not assumed. In the inner, optically thick layers of the supernova envelope where thermalization takes place, the diffusion of equilibrium radiation is described in the approximation of radiative heat conduction. We have performed our study with a simplified version of radiation hydrodynamics in the one-group (gray) approximation, in which the nonequilibrium radiation field can be parametrized by an appropriate blackbody temperature.

### Radiation Hydrodynamics in the One-Group Approximation

According to Mihalas and Mihalas (1984),[1] the system of radiation hydrodynamics equations in the

---

[1] See also an independent derivation of the ensuing equations by Imshennik (1993) based on the book by Imshennik and Morozov (1981).

one-group approximation comprises the following: the continuity equation

$$\frac{\partial r}{\partial t} = u, \quad \frac{\partial r}{\partial m} = \frac{1}{4\pi r^2 \rho}, \quad (1)$$

the equation of motion

$$\frac{\partial u}{\partial t} = -4\pi r^2 \frac{\partial (P_g + Q)}{\partial m} - \frac{Gm}{r^2} + \frac{1}{c}\chi_F^0 F^0, \quad (2)$$

the energy equation for the gas

$$\frac{\partial E_g}{\partial t} = -(P_g + Q)\frac{\partial}{\partial t}\left(\frac{1}{\rho}\right) + c\kappa_E^0 E^0 - 4\pi \frac{\eta_t^0}{\rho} + \varepsilon, \quad (3)$$

the equation for the total radiative energy density

$$\frac{\partial E^0}{\partial t} = -4\pi\rho \frac{\partial}{\partial m}(r^2 F^0) - 4\pi\rho(1 + f^0) \quad (4)$$
$$\times E^0 \frac{\partial}{\partial m}(r^2 u) + \frac{u}{r}(3f^0 - 1)E^0 + 4\pi\eta_t^0 - c\rho\kappa_E^0 E^0,$$

and the equation for the total radiative energy flux

$$\frac{\partial F^0}{\partial t} = \left(2\frac{u}{r} - c\rho\chi_F^0 - 8\pi\rho\frac{\partial}{\partial m}(r^2 u)\right)F^0 \quad (5)$$
$$- c^2\left(4\pi r^2 \rho \frac{\partial}{\partial m}(f^0 E^0) + \frac{1}{r}(3f^0 - 1)E^0\right).$$

Here, $t$ is the time in the comoving frame of reference; $m$ is the mass of the gas within the sphere of radius $r$; $u$ is the gas velocity; $\rho$ is the gas density; $P_g$ and $E_g$ are the pressure and specific internal energy of the gas, respectively; $Q$ is the artificial viscosity; $E^0$ is the total radiative energy density related to the radiation temperature $T_r$ by $E^0 = aT_r^4$; $F^0$ is the total radiative energy flux; $\chi_F^0$ is the mean opacity weighted in radiative energy flux (true absorption and scattering); $\kappa_E^0$ is the mean opacity weighted in radiative energy density (true absorption); $\eta_t^0$ is the total thermal emission coefficient; $\varepsilon$ is the rate of change in internal energy, for example, due to the deposition of gamma rays produced by radioactive decays; $f^0$ is the variable Eddington factor equal to the ratio of the radiation pressure $P_r^0$ to the total radiative energy density $E^0$. The radiation characteristics, the mean opacities, and the thermal emission coefficient refer to the comoving frame of reference and are denoted by the superscript 0.

Initial data and boundary conditions must be specified to properly formulate the problem of radiation hydrodynamics described by hyperbolic equations: the radius $r$, the velocity $u$, and the total radiative energy flux $F^0$ at the inner boundary and the total radiative energy flux $F^0$ and a zero gas pressure ($P_g = 0$) at the outer boundary.

*Ionization Balance and the Equation of State*

If the stationarity condition is satisfied, the equation of state for an ideal gas in a nonequilibrium radiation field and for nonthermal ionization requires solving the problem of the populations of excited atomic and ionic levels and the ionization balance. In the absence of LTE, the distribution in excited atomic and ionic levels and in ionization states is determined by the balance of all possible elementary processes, and the general solution of the problem must include an infinite system of algebraic equations for the entire set of levels and processes. This is a very complicated problem because of both the mathematical difficulties and the lack of reliable physical data on the required elementary processes. This circumstance and the necessity of multiple calculations of the equation of state in hydrodynamic calculations force us to disregard the excited atomic and ionic levels and to restrict our analysis only to the ground atomic and ionic states and, hence, to their ionization balance. Of greatest interest in the establishment of ionization balance in supernova envelopes are the following elementary processes: photoionization and radiative recombination, electron ionization, three-particle recombination, and nonthermal ionization.

To simplify the system of balance equations, we take into account only three ionization states for all elements. The ionization balance equation for a $Z^0$ atom and a $Z^+$ ion in which the photoionization, electron ionization, and nonthermal ionization rates are balanced out by the radiative and three-particle recombination rates is

$$R_{Z^0} N_{Z^0} + q_{Z^0} N_e N_{Z^0} + \Gamma_{Z^0} N_{Z^0} \quad (6)$$
$$= \alpha_{Z^+} N_e N_{Z^+} + \chi_{Z^+} N_e^2 N_{Z^+},$$

where $R_{Z^0}$ is the photoionization probability of the $Z^0$ atom in a radiation field that is not necessarily in equilibrium; $\alpha_{Z^+}$ is the radiative recombination coefficient; $q_{Z^0}$ is the electron ionization rate for the $Z^0$ atom for a Maxwellian energy distribution; $\chi_{Z^+}$ is the coefficient of the corresponding three-particle recombination; $\Gamma_{Z^0}$ is the nonthermal ionization rate for the $Z^0$ atom; and $N_e$, $N_{Z^0}$, and $N_{Z^+}$ are the number densities of the electrons, $Z^0$ atoms, and $Z^+$ ions, respectively. A similar equation can also be written for the $Z^+$ and $Z^{++}$ ions and for hydrogen, the only difference being that the negative hydrogen ion will act as the third ionization state.

The analytic fits to the cross sections for photoionization from the ground state for various atoms and ions required to calculate the photoionization probability and the radiative recombination coefficient were taken from the papers by Verner and Yakovlev (1995) and Verner *et al.* (1996). The photoionization cross section for the negative hydrogen ion was calculated by Wishart (1979). The fitting formulas for the electron ionization rates of various atoms and ions are given in the paper by Voronov (1997). The radioactive nuclides produced by explosive nucleosynthesis emit mostly gamma rays with an energy of about 1 MeV, which lose their energy through Compton scattering by free and bound electrons. In turn, the Compton electrons lose their energy through the heating of free electrons and the ionization and excitation of atoms and ions. Kozma and Fransson (1992) accurately solved this complicated problem and calculated the nonthermal ionization and excitation rates for atoms and ions.

Now, to find the ionization balance for a mixture of chemical elements, for example, composed of H and elements from He to Fe, we close the system of balance equations described above by the equations for the conservation of the number of particles and electrical plasma neutrality. Given the gas density, the electron temperature, the nonequilibrium radiation and nonthermal ionization properties, and the parameters of the elementary processes, the derived system of equations can be easily reduced to a nonlinear equation for the unknown degree of ionization $x_e$. Note that the physical conditions in the supernova envelope at the expansion phase concerned are characterized by the equality between the electron and ion temperatures, which are designated as the gas temperature $T_g$. Solving this equation for the degree of ionization $x_e$ yields all of the required relative atomic and ionic number densities $x_{H^0}$, $x_{H^+}$, $x_{H^-}$, $x_{Z^0}$, $x_{Z^+}$, and $x_{Z^{++}}$.

The pressure $P_g$ and the internal energy $E_g$ of an ideal gas in the equation of state for a mixture of chemical elements are

$$P_g = \frac{kT_g}{m_u A}(1 + A x_e)\rho, \qquad (7)$$

$$E_g = \frac{3}{2}\frac{kT_g}{m_u A}(1 + A x_e) \qquad (8)$$
$$+ \frac{X_H}{m_u A_H}(I_H x_{H^+} - I_{H^-} x_{H^-})$$
$$+ \sum_{Z=He}^{Fe} \frac{X_Z}{m_u A_Z}\left(I_{Z^0} x_{Z^+} + \left(I_{Z^0} + I_{Z^+}\right) x_{Z^{++}}\right),$$

where $X_Z$ is the mass fraction of the Z element; $A_Z$ is the atomic mass of the Z element (AMU); $I_{Z^0}$ and $I_{Z^+}$ are the ionization potentials for the $Z^0$ atom and the $Z^+$ ion, respectively; and $A$ is the mean atomic mass (AMU) defined by the identity

$$\frac{1}{A} = \frac{X_H}{A_H} + \sum_{Z=He}^{Fe} \frac{X_Z}{A_Z}. \qquad (9)$$

The ionization balance and the equation of state were considered in more detail previously (Utrobin 1998).

### The Mean Opacities and the Emission Coefficient

The relative atomic and ionic number densities that were calculated, as described above, in the absence of LTE, but without allowing for any excited states, determine the corresponding mean opacities and the thermal emission coefficient. As the mean opacity $\chi_F^0$ weighted in radiative energy flux, we use the Rosseland mean that includes the contributions both from bound–free and free–free absorption and from Thomson scattering by free electrons and Rayleigh scattering by neutral hydrogen. The mean opacity $\kappa_E^0$ weighted in radiative energy density is calculated as the Planck mean with a radiation temperature $T_r$ that includes only the contributions from bound–free and free–free absorption. Accordingly, the total spontaneous thermal emission coefficient $\eta_t^0$ includes bound–free and free–free processes. The free–free absorption by atoms and ions was calculated using the Gaunt factor (tabulated by Sutherland 1998) averaged over the Maxwellian distribution, and the free–free absorption coefficient for the negative hydrogen ion was taken from the paper by Bell and Berrington (1987).

Apart from the processes in the continuum, we also took into account the contribution of bound–bound processes to the opacity. Slightly more than 500 000 spectral lines were chosen from the extensive atomic line database compiled by Kurucz and Bell (1995). To calculate the corresponding opacity, we solved the special problem of the populations of excited atomic and ionic levels and the ionization balance for a mixture of chemical elements from H to Zn with all ionization stages at given density and temperature using the Boltzmann and Saha formulas. We averaged the contribution of spectral lines in a medium with a velocity gradient using the generalized formula by Castor *et al.* (1975) at the early expansion phases of the supernova envelope and using the formula by Blinnikov (1996) after the passage to free expansion. The line opacity in an expanding medium calculated in this way was treated as pure scattering.

### The Numerical Method

We used the standard method of lines (see, e.g., Berezin and Zhidkov 1962; Blinnikov and Bartunov 1993) to numerically solve the system of radiation hydrodynamics equations (1)–(5). The essence of this method is that the system of partial differential equations is reduced to a system of ordinary differential equations by an appropriate substitution of finite-difference fits for the derivatives with respect to the spatial coordinates. In all of the computed models, we broke the star down into 300 computational mass bins. The derived system of ordinary differential equations is stiff, but there are well-developed algorithms for its integration. One of the most efficient methods is the implicit method by Gear (1971) with an automatic choice of both the time integration step and the order of accuracy of the method. The implementation of Gear's method involves calculating the corresponding Jacobi matrix and inverting the large and sparse matrix derived from it. Zlatev's efficient algorithm (Österby and Zlatev 1983) that was specially modified by Blinnikov and Bartunov (1993) was used to invert such a large and sparse matrix of arbitrary structure.

The system of time-dependent frequency-integrated moment equations—the equation for the total radiative energy density (4) and the equation for the total radiative energy flux (5)—is closed by introducing a variable Eddington factor $f^0$. The latter can be determined by solving the time-independent transfer equation for a spherically symmetric expanding medium (Mihalas and Mihalas 1984) written simultaneously in the laboratory and comoving frames of reference to within terms on the order of $u/c$,

$$\mu \frac{\partial I}{\partial r} + \frac{1-\mu^2}{r}\frac{\partial I}{\partial \mu} = \left(1 - \mu\frac{u}{c}\right)\rho(\kappa_a^0 + \kappa_s^0)(S - I), \quad (10)$$

with the source function that explicitly includes the contribution of scattering,

$$S = \left(1 + 5\mu\frac{u}{c}\right)\frac{\eta_t^0/\rho + \kappa_s^0 J^0}{\kappa_a^0 + \kappa_s^0}, \quad (11)$$

where $\mu = \cos\theta$ and $\theta$ is the angle between the direction of propagation of the radiation and the radius vector $\mathbf{r}$. The frequency-integrated specific radiation intensity $I$ refers to the laboratory frame of reference, while the frequency-integrated mean radiation intensity $J^0$, the opacity of true absorption $\kappa_a^0$, and the opacity of scattering $\kappa_s^0$ are calculated in the comoving frame of reference and are denoted by the superscript 0. The formal solution along the characteristics (individual rays) in the absence of externally incident radiation on the outer boundary is used in solving the transfer equation (10). The frequency-integrated moments of the radiation intensity are transformed from the laboratory frame to the comoving frame using the Lorenz transformations (Mihalas and Mihalas 1984), and the corresponding Eddington factor $f^0$ is then calculated. In addition, the solution obtained allows us to also calculate the variable Eddington factor at the outer boundary, which relates the total radiative energy flux to the mean frequency-integrated radiation intensity.

The bolometric luminosity calculated by taking into account the delay and limb-darkening effects is given by the formula

$$L(t_{\text{obs}}) = \int_0^1 4\pi R^2(t) \pi I(R(t), \mu) 2\mu d\mu, \quad (12)$$

where $t_{\text{obs}}$ is the time in the observer's frame of reference; $R(t)$ is the outer radius of the supernova envelope at time $t$; $I(R(t), \mu)$ is the specific radiation intensity at the outer boundary of the envelope at the same time derived from the solution of the transfer equation (10) with the angular dependence and with the appropriate source function (11). It would be natural to measure the time $t_{\text{obs}}$ from the detection time of the neutrino signal that emerges during the supernova explosion and that reaches an observer at distance $D$ in time $D/c$. In this case, the relation between the times $t_{\text{obs}}$ and $t$ is given by

$$t_{\text{obs}} = t - \frac{R(t)\mu}{c}. \quad (13)$$

The deposition of gamma rays produced by radioactive $^{56}$Ni and $^{56}$Co decays can be determined from the solution of the gamma-ray transfer problem by assuming that the gamma rays interact with matter through absorption at opacity $\kappa_\gamma = 0.06 Y_e$ cm$^2$ g$^{-1}$, where $Y_e$ is the number of electrons per baryon. The gamma-ray transfer was modeled by the transfer equation (10) with appropriate opacity $\kappa_\gamma$ and source function.

HYDRODYNAMIC MODELS

First, recall that the presupernova is Sanduleak $-69°202$, a blue B3 Ia supergiant (Gilmozzi et al. 1987; Panagia et al. 1987; Sonneborn et al. 1987) with an apparent magnitude of $12^m\!24$ (Rousseau et al. 1978), an effective temperature of about 16 300 K, and a bolometric correction of about $-1.15$ (Humphreys and McElroy 1984). At the distance modulus $m - M = 18.5$ for the LMC, the color excess $E(B-V) = 0.15$, and the interstellar extinction $A_V = 3.1E(B-V)$ (Pun et al. 1995), the radius of the blue supergiant is about $46.8 R_\odot$. We take this value as the presupernova radius $R_0$.

Important information about the chemical composition of the surface layers in the presupernova can be obtained by analyzing the spectra of its circumstellar matter. Studying emission lines from the circumstellar matter around SN 1987A, Wang (1991) estimated the ratio $n(\text{He})/n(\text{H}) \sim 0.20$. Investigating narrow ultraviolet and optical emission lines from the inner circumstellar ring around SN 1987A, Lundqvist and Fransson (1996) obtained $n(\text{He})/n(\text{H}) \sim 0.25$. Being aware of the uncertainty in these estimates, we take a ratio of 0.20. This value together with the data by Dufour (1984) on the metallicity in the LMC yields a relative mass fraction of $X = 0.555$ for hydrogen, $Y = 0.441$ for helium, and $Z = 0.004$ for heavy elements. We take this chemical composition as the chemical composition of the surface layers in the presupernova. According to Dufour (1984), a chemical composition with $X = 0.743$, $Y = 0.251$, and $Z = 0.006$ is characteristic of the LMC.

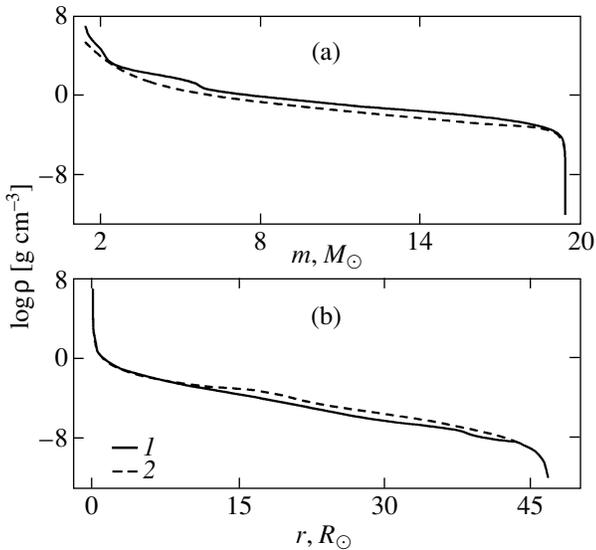

**Fig. 1.** Initial density distributions in mass (a) and radius (b) for the evolutionary model E (*1*) and the nonevolutionary model N (*2*).

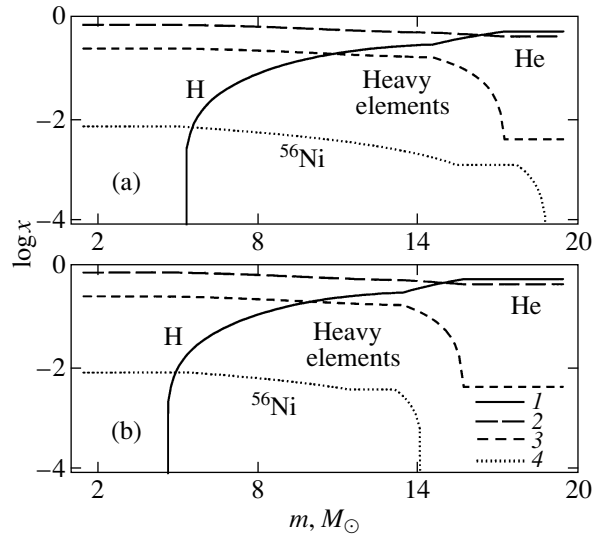

**Fig. 2.** Chemical composition of the evolutionary model E (a) and the nonevolutionary model N (b): *1*—hydrogen mass fraction, *2*—helium mass fraction, *3*—heavy-element mass fraction, *4*—$^{56}$Ni mass fraction.

Here, we investigate the evolutionary and nonevolutionary models of the presupernova. The evolutionary models were constructed from model l20n2ae that was computed by Woosley *et al.* (1997) without allowing for mass loss and rotation and with a collapsing core mass of $1.58 M_\odot$, a helium core mass of $5.85 M_\odot$, an ejected envelope mass of $17.80 M_\odot$, a total mass of $19.38 M_\odot$, and an outer radius of $64.2 R_\odot$. For convenience and agreement with observations of the presupernova, model l20n2ae was rescaled to a hydrostatic presupernova model with the ejected envelope mass $M_{\text{env}} = 18.0 M_\odot$, the total mass $M = 19.58 M_\odot$, and the initial radius $R_0 = 46.8 R_\odot$; the outer layers were broken down into smaller computational mass bins that were in both hydrostatic and radiative equilibrium at the observed presupernova bolometric luminosity of $-8^m\!.12$. Strictly speaking, after this restructuring and after specifying a chemical composition of the outer layers identical to the observed presupernova composition, the model can hardly be called evolutionary. Nevertheless, a very important property of the evolutionary model l20n2ae, the pattern of density distribution at the explosion time (Fig. 1), is preserved. Below, we denote these models by the letter E.

The nonevolutionary presupernova models have the same ejected envelope mass $M_{\text{env}}$, total mass $M$, and initial radius $R_0$ as the evolutionary models, but differ from them in initial density distribution (Fig. 1). The latter is chosen from the condition of good agreement between the computed and observed bolometric light curves for moderate $^{56}$Ni mixing, by analogy with the construction of the previous hydrodynamic model for SN 1987A (Utrobin 1993). Just as in the evolutionary presupernova models, the outer layers of the nonevolutionary models are in hydrostatic and radiative equilibrium at the observed presupernova bolometric luminosity and have the observed chemical composition of the presupernova, unless specified otherwise, while the inner layers have the chemical composition of model l20n2ae (Fig. 2). We denote these models by the letter N.

The supernova explosion is triggered by an instantaneous energy release near the stellar center at the initial time. The explosion energy $E$ is specified as the excess above the total energy of the initial envelope configuration. In all of the models under consideration, except for the specially specified ones, the mass of the radioactive nickel nuclide $M_{\text{Ni}}$ is $0.073 M_\odot$. Basic parameters of the computed hydrodynamic models are given in the table. Apart from the basic models E and N and the similar models Enn

Parameters of the hydrodynamic models

| Model | $M_{env}$, $M_\odot$ | $R_0$, $R_\odot$ | $E$, $10^{51}$ erg | $M_{Ni}$, $M_\odot$ | Note |
|---|---|---|---|---|---|
| E   | 18 | 46.8 | 1.0 | 0.073 | – |
| Eni | 18 | 46.8 | 1.0 | 0.073 | $^{56}$Ni profile, as in model N |
| Enn | 18 | 46.8 | 1.0 | 0     | Without $^{56}$Ni |
| N   | 18 | 46.8 | 1.0 | 0.073 | – |
| Nnn | 18 | 46.8 | 1.0 | 0     | Without $^{56}$Ni |
| Nnt | 18 | 46.8 | 1.0 | 0.073 | Without nonthermal ionization |
| Nlo | 18 | 46.8 | 1.0 | 0.073 | Without line opacity |
| Nld | 18 | 46.8 | 1.0 | 0.073 | Without limb darkening |
| Ncc | 18 | 46.8 | 1.0 | 0.073 | LMC chemical composition |
| Nce | 18 | 46.8 | 1.2 | 0.073 | LMC chemical composition |

and Nnn, but without $^{56}$Ni, we present the following: model Eni shows moderate $^{56}$Ni mixing in the evolutionary model; model Nnt was computed without including nonthermal ionization; model Nlo shows the role of line opacity in the expanding medium; model Nld indicates the importance of limb darkening in calculating the emergent flux from the supernova envelope; models Ncc and Nce have a chemical composition of the surface layers characteristic of the LMC.

## RESULTS

### The Evolutionary and Nonevolutionary Presupernova Models

Woosley (1988), Shigeyama and Nomoto (1990), and Blinnikov et al. (2000) have convincingly shown that in the explosion of the presupernova model obtained through evolutionary calculations, good agreement with the observed bolometric light curve is achieved only under the assumption of strong $^{56}$Ni mixing up to velocities of ∼4000 km s$^{-1}$. Indeed, moderate (at velocities $u \leq 2500$ km s$^{-1}$) $^{56}$Ni mixing in the evolutionary model Eni similar to the mixing in the nonevolutionary model N (Fig. 3) gives rise to a local minimum in the light curve in the interval 15–30 days (Fig. 4). In the evolutionary model, this deficit of bolometric luminosity, as was pointed out above, can be removed only through strong $^{56}$Ni mixing, which is demonstrated by model E (Figs. 3 and 4).

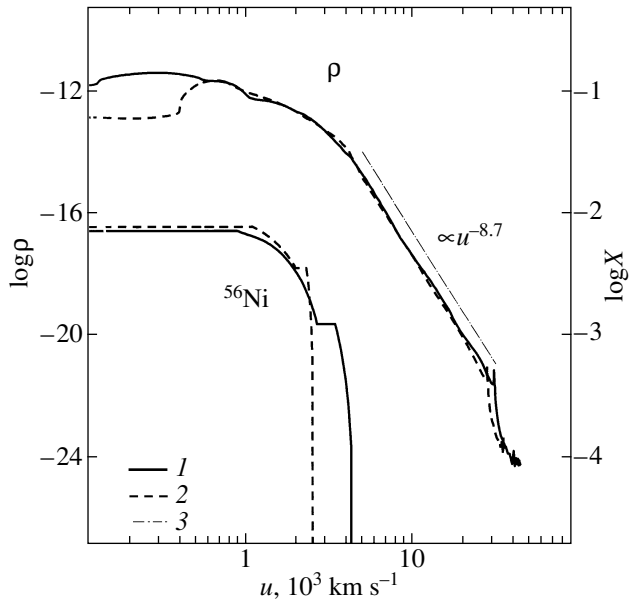

**Fig. 3.** Behavior of the density and the $^{56}$Ni mass fraction in the supernova envelope at time $t = 120$ days: (*1*) model E, (*2*) model N, (*3*) density distribution fit $\rho \propto u^{-8.7}$ in the velocity range 4400–27 000 km s$^{-1}$.

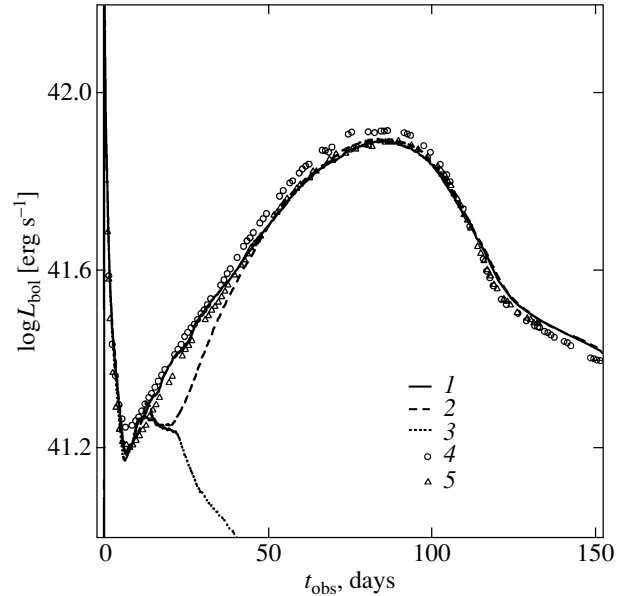

**Fig. 4.** Bolometric light curves for model E (*1*), model Eni (*2*) (model E with the $^{56}$Ni profile in velocity space as in model N), model Enn (*3*) (model E without $^{56}$Ni), the observations of SN 1987A (*4*) by Catchpole (1987, 1988), the observational data (*5*) by Hamuy et al. (1988).

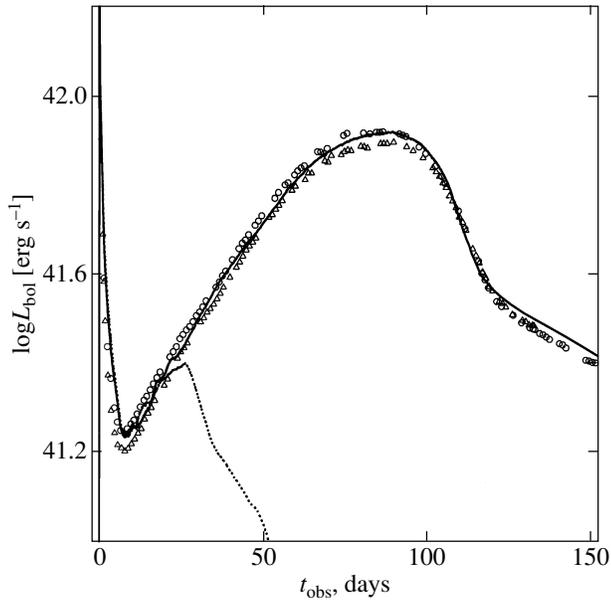

**Fig. 5.** Same as Fig. 4 for model N (solid line) and model Nnn—model N without $^{56}$Ni (dotted line).

The local minimum of the bolometric luminosity in the evolutionary model Eni arises as follows: Energy release in the central regions of the presupernova gives rise to a strong shock that propagates toward the stellar surface. After the shock passage through the star, the matter heats up, and the radiative energy density increases. Concurrently, the matter acquires an outward-increasing velocity that everywhere exceeds the local parabolic velocity. In about 0.0823 days, the shock emerges on the surface and heats the outer stellar layers—the effective temperature rises to ∼4.6×10$^5$ K, the color temperature rises to ∼1.1 ×10$^6$ K, and the luminosity increases sharply. After the emergence of the shock, an expansion of the star begins; this expansion leads to rapid cooling of the surface layers and to the same rapid decrease in luminosity. As a result, a narrow peak is formed whose luminosity at maximum is ∼3.5 ×10$^{44}$ erg s$^{-1}$ and which is partially shown in the bolometric light curve (Fig. 4).

The subsequent expansion of the envelope gradually creates conditions under which a special radiative cooling regime of the matter arises—a cooling wave (CW). Remarkably, even the first hydrodynamic models of supernovae constructed and studied by Imshennik and Nadyozhin (1964) revealed this cooling regime for compact presupernovae, to which the case of SN 1987A belongs. The CW properties were studied in more detail by Grassberg and Nadyozhin (1976). After about 10 days, this regime for the ejected matter completely determines the supernova luminosity; the photosphere is located within the CW front, and the subphotospheric, optically thick layers are cooled almost adiabatically. In the absence of $^{56}$Ni in the supernova envelope, the internal energy of the matter and the radiative energy stored in these layers after the shock passage and then partially lost during expansion continue to radiate away for a relatively short period. In the evolutionary model Enn, which is similar to E and which contains no $^{56}$Ni in the envelope, the supernova luminosity begins to decrease already after ∼13 days, and the energy store in the envelope is depleted by ∼40 days (Fig. 4).

Clearly, if $^{56}$Ni is distributed in the deep layers whose time of gamma-ray diffusion remains larger than the characteristic envelope expansion time for a long period, then the gamma-ray diffusion begins to be involved in shaping the light curve with a long delay when the luminosity already begins to fall; as a result, a local minimum of the bolometric luminosity is formed. Moderate $^{56}$Ni mixing in the evolutionary model Eni has the property described above (Fig. 4). To obtain a monotonic light curve without any local minimum of the bolometric luminosity in the interval 15–30 days, the following simple solution suggests itself: strong $^{56}$Ni mixing up to velocities of ∼4000 km s$^{-1}$ at which the local gamma-ray deposition compensates for the energy loss through envelope expansion and prevents the undesirable decrease in luminosity. This effect, which is demonstrated by model E (Fig. 4), helped Woosley (1988), Shigeyama and Nomoto (1990), and Blinnikov *et al.* (2000) to reproduce the observed bolometric light curve in their hydrodynamic calculations.

A more radical method of obtaining a monotonic light curve even at moderate $^{56}$Ni mixing is to construct a model of the presupernova with a density of its outer layers much higher than the density in the evolutionary model l20n2ae. In such a presupernova, more internal energy of the matter and more radiative energy is stored in the outer, denser layers after the passage of a strong shock through the envelope; this energy may prove to be enough to maintain the increasing luminosity even after 13 days. A proper density distribution in the now nonevolutionary presupernova model (denoted by the letter N) is obtained from agreement between the theoretical and observed bolometric light curves; the density ratio in the outer ($15-40R_\odot$) layers lies within the range 2.4–4.9 (Fig. 1). Indeed, this density distribution in the presupernova even without $^{56}$Ni in the envelope causes an increase in luminosity until ~25 days, as shown by model Nnn (Fig. 5). Finally, moderate $^{56}$Ni mixing in model N (Fig. 3) proves to be enough to explain the observed behavior of the bolometric light curve (Fig. 5), just as in our previous hydrodynamic model (Utrobin 1993).

Both in the evolutionary model E after ~13 days (Fig. 4) and in the nonevolutionary model N after ~25 days (Fig. 5), the subsequent run of the bolometric light curve is entirely determined by gamma-ray energy deposition and by the subsequent reprocessing of this energy by the envelope matter and its reradiation in the form of soft thermal photons. The ascending branch of the light curve and its peak were reconciled with observations by an appropriate specification of the chemical composition in the transition region between the surface layers and the helium core (Fig. 2). Here, the crucial factor is deep hydrogen mixing into the helium core up to velocities of ~980 km s$^{-1}$. This role of hydrogen mixing in reproducing the bolometric light curve was previously noted by Woosley (1988), Shigeyama and Nomoto (1990), Utrobin (1993), and Blinnikov et al. (2000). The quasi-exponential decline in luminosity after the maximum is caused by radioactive decays of $^{56}$Ni and $^{56}$Co whose mass is $0.073 M_\odot$.

*The Gas and Radiation Temperature Profiles*

According to Eqs. (3) and (4), abandoning LTE when solving the equation of state and determining the mean opacities and the thermal emission coefficient will affect the pattern of energy exchange between matter and radiation. The results of this approach can be best studied from the behavior of the gas temperature and the radiation temperature in the envelope. Using the main model N as an example, let us trace the evolution of the radial distributions of the gas temperature and the radiation temperature in the comoving frame of reference at the most characteristic times shown in Fig. 6. Clearly, from the center to the subphotospheric layers in an optically thick medium under conditions close to LTE, the radiation temperature is virtually equal to the gas temperature (Figs. 6a–6c). This state exists until $t \sim 115$ days. Subsequently, however, the envelope rapidly becomes optically thin, and the gas temperature turns out to be lower than the radiation temperature because of the weakening interaction between matter and radiation against the background of the increasing role of adiabatic losses (Fig. 6d).

In contrast, the gas temperature above the photosphere differs from the radiation temperature even in the initial state due to non-LTE conditions. Subsequently, when the shock emerges on the surface and several days later, this difference gradually increases and then becomes significant (Fig. 6). The bolometric luminosity in the surface layers in the comoving frame of reference is almost constant: $L^0 = 4\pi r^2 \pi F^0 \approx$ const (Fig. 7). This fact may be expressed in the following approximate dependence for the radiation temperature distribution in the outer layers: $T_r \propto (E^0)^{1/4} \propto (F^0)^{1/4} \propto r^{-1/2}$. However, the high expansion velocity of these layers and the delay effect lead to a deviation from this simple dependence. During the explosion, the bolometric luminosity decreases outward; it "remembers" the earlier and bright phase only after the liminosity peak in the interval $t \approx 120-140$ days and increases toward the envelope surface (Fig. 7). The described behavior of the bolometric luminosity is consistent with the calculations by Blinnikov et al. (2000). Accordingly, during the explosion, the radiation temperature curve runs below the fitting straight line (Figs. 6a–6c), except for the interval $t \approx 120-140$ days when the radiation temperature is higher (Fig. 6d).

In turn, the gas temperature in the surface layers in the almost complete absence of interaction between radiation and matter and of any heating sources decreases adiabatically as $T_g \propto t^{-2}$ (Fig. 6) even since the first days. A "frozen-in" structure, a temperature trace from the thin, dense layer formed when the shock emerges on the presupernova surface, arises at the outer edge of the envelope. The dense layer itself is produced by the transition from the adiabatic regime of propagation of a strong shock in the surface layers to the isothermal regime. This layer was first observed in hydrodynamic models of SN II with extended progenitors similar to supergiants (Grassberg et al. 1971). For SN 1987A, Imshennik and Nadyozhin (1988) analytically considered the emergence of a shock wave on the presupernova surface. In the hydrodynamic modeling of SN 1987A, a dense layer is formed only when abandoning the approximation of radiative heat conduction, which is too rough to describe the radiative transfer in the surface layers (Blinnikov and Nadyozhin 1991; Ensman and Burrows 1992; Blinnikov et al. 2000). The temperature trace is subsequently observed all of the time (it is not shown in Figs. 6c and 6d). Until $t \sim 115$ days, a transition from the optically thick medium where the gas and radiation are described by a single temperature to the transparent layers where the gas cools down adiabatically occurs in the region between the photosphere and the surface layers. In this case, the complex interaction between the heating and cooling processes in the near-photosphere layer manifests itself in the form of a thin structure in the gas temperature distribution (Figs. 6a, 6c, and 6d).

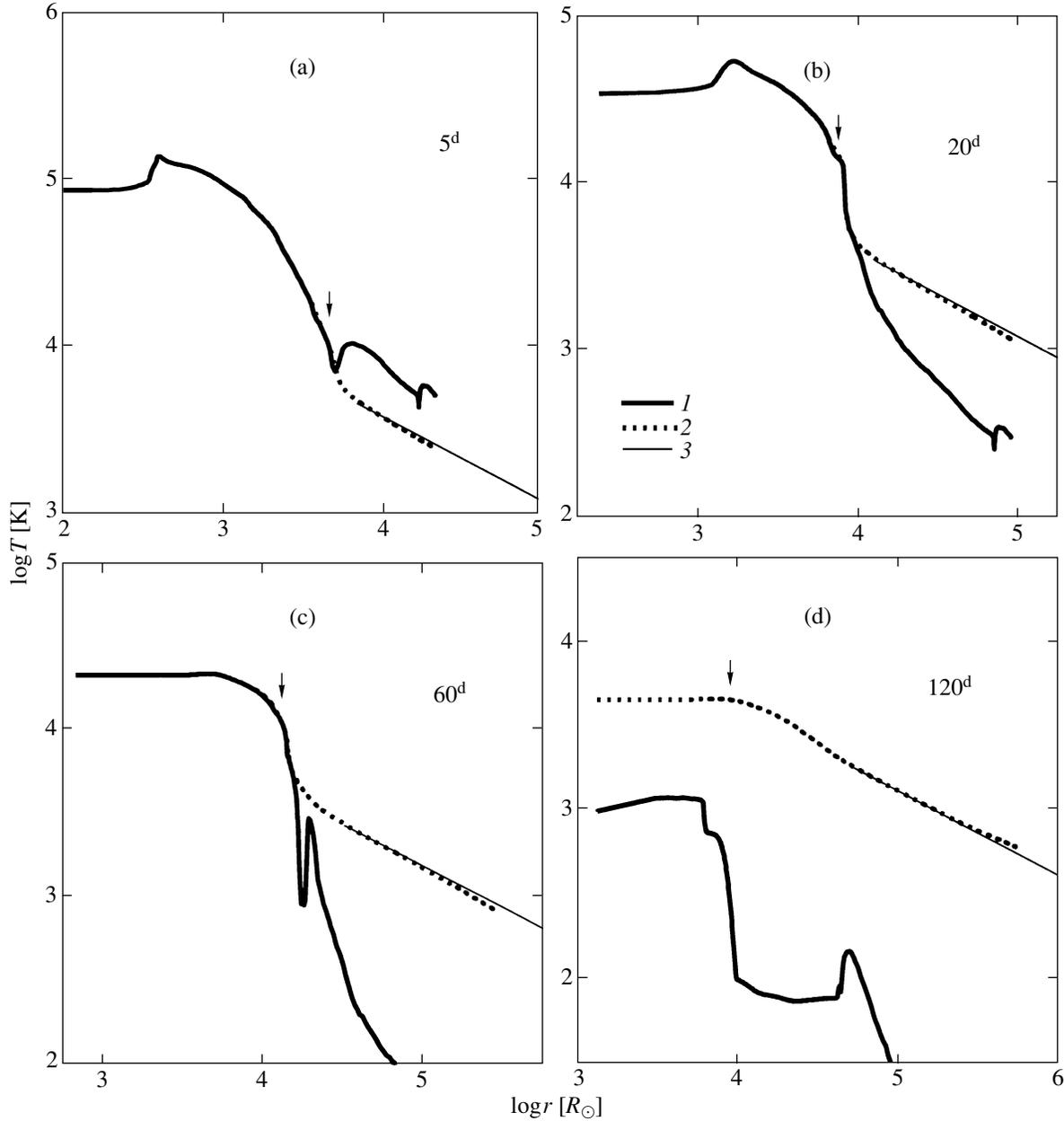

**Fig. 6.** Evolution of the radial distribution of the gas temperature (*1*) and the radiation temperature (*2*) in the comoving frame of reference in model N: (a) $t = 5$ days, (b) $t = 20$ days, (c) $t = 60$ days, (d) $t = 120$ days; *3*—the fit $T_r \propto r^{-1/2}$ to the radiation temperature distribution under the assumption of constant bolometric luminosity. The vertical arrow marks the location of the photosphere.

### The Role of Nonthermal Ionization

After $t_{\rm obs} \sim 30$ days, the luminosity of SN 1987A is produced mainly by $^{56}$Ni and $^{56}$Co decays, and the pertinent question regarding the role of nonthermal ionization in the supernova explosion arises. To study this question, we computed model Nnt, in which the ionization balance was found without allowing for nonthermal ionization in Eq. (6), but with the energy deposition from radioactive $^{56}$Ni and $^{56}$Co decays retained in Eq. (3). The bolometric light curve for model Nnt turned out to be almost coincident with that for model N, exhibiting only a small luminosity deficit as the maximum is approached in the interval $t_{\rm obs} \approx 40\text{–}70$ days (Fig. 8). This deficit owes its origin to an increase in opacity in the absence of nonthermal ionization. Thus, it should be recognized that nonthermal ionization plays a minor role in shaping the bolometric light curve of SN 1987A. A much larger effect of nonthermal ionization was found in SN 1993J (Utrobin 1996). This finding can be explained by the much lower hydrogen abundance in the SN 1993J envelope than that of the SN 1987A envelope and, as a result, by the larger relative contribution of other elements to the opacity when nonthermal ionization is taken into account.

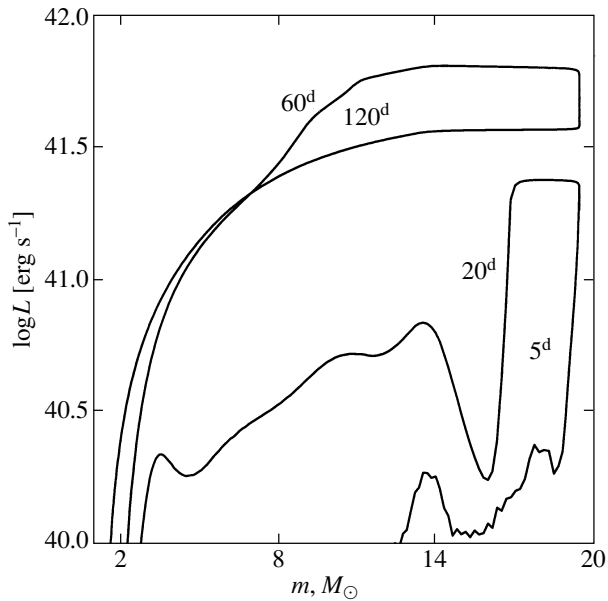

**Fig. 7.** Luminosity distribution in mass in the comoving frame of reference for model N at times $t = 5$, 20, 60, and 120 days.

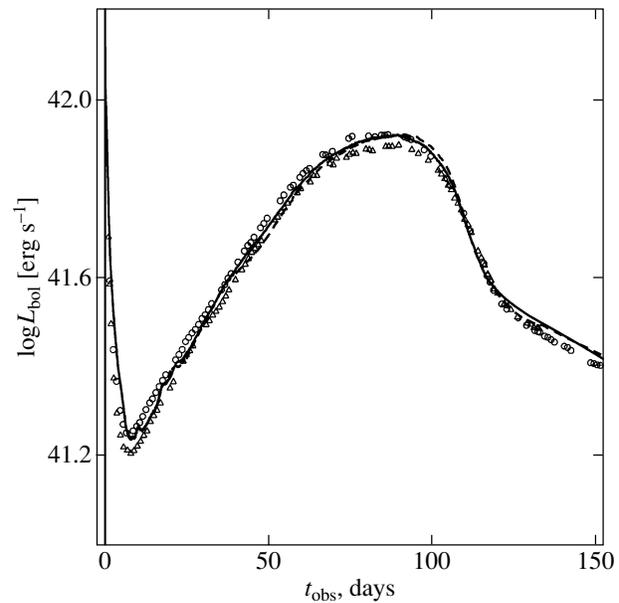

**Fig. 8.** Same as Fig. 4 for model N (solid line) and model Nnt—model N without nonthermal ionization (dashed line).

Before the envelope becomes optically thin, the behavior of the gas temperature and the radiation temperature with nonthermal ionization (Fig. 6) differs little from the case without nonthermal ionization (Fig. 9). However, at $t = 120$ days, a radical difference arises in the inner layers in which the main energy deposition from radioactive $^{56}$Co decays takes place: the gas temperature is lower than the radiation temperature in the former case and vice versa in the latter. The enhanced ionization of the matter caused by nonthermal ionization increases the importance of bremsstrahlung processes in the energy exchange between matter and radiation and even leads to the dominance of these cooling processes over the nonthermal gas heating. In the absence of nonthermal ionization, the gas cooling by bremsstrahlung processes becomes much weaker, and the nonthermal heating dominates, causing the gas temperature to exceed the radiation temperature.

### The Influence of Line Opacity

The decrease in opacity due to disregarding the line contribution in model Nlo causes the bolometric luminosity to increase from the first few days and until the luminosity peak (Fig. 10), which is equivalent in action to a decrease in opacity due to the reduction in hydrogen abundance (Utrobin 1989). The corresponding behavior of the light curve also shows up when increasing the hydrogen abundance in the outer layers (Fig. 11). These facts lead us to the firm conclusion that disregarding the contribution of lines to the opacity introduces an error of ~20% into the explosion energy $E$ (see the table). Since an increase in explosion energy $E$ by a factor of 1.5 at the CW phase increases the bolometric luminosity by ~1.5 and since an increase in envelope mass $M_{env}$ by the same factor decreases the luminosity by a factor of ~1.3 (Utrobin 1989), a similar error is also possible when determining the mass of the ejected matter. Recall that this dependence of the bolometric luminosity at a fixed initial radius of the presupernova makes it difficult to estimate the explosion energy $E$ and the ejected envelope mass $M_{env}$, and the observed luminosity actually determines only their ratio $E/M_{env}$ (Woosley 1988; Utrobin 1993).

After the luminosity peak, the bolometric luminosity in model Nlo decreases more rapidly than the observed luminosity, which emphasizes the fundamental role of the line contribution to opacity, along with the proper description of the radiative transfer, in reproducing the observed shape of the bolometric light curve (Fig. 10). Due to the disregard of the line contribution to the opacity and without a proper allowance for the radiative transfer, we failed to achieve reasonable agreement with the observed light curve near and after the peak in the hydrodynamic modeling

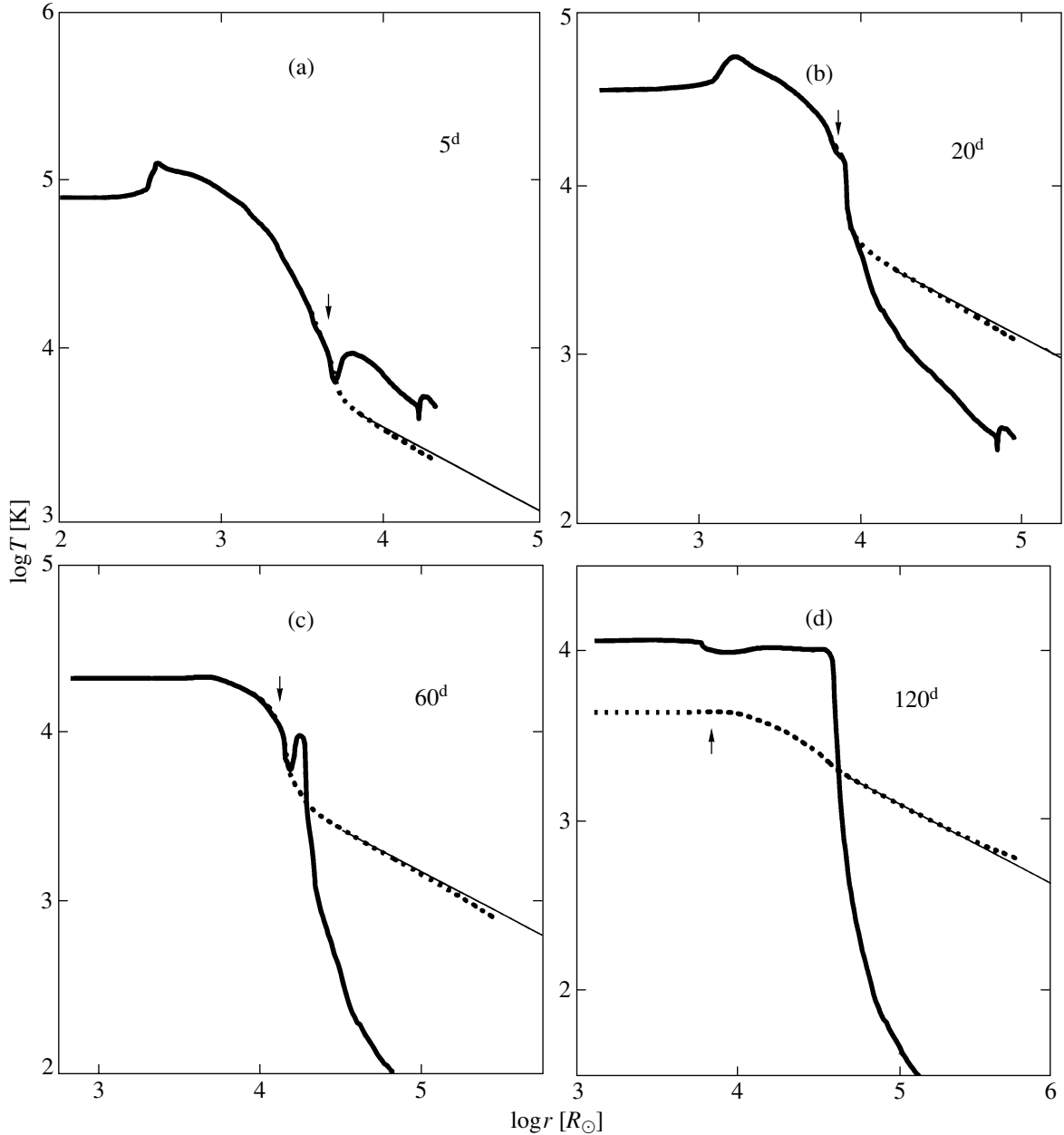

**Fig. 9.** Same as Fig. 6 for model Nnt.

of SN 1987A using the approximation of radiative heat conduction (Utrobin 1993).

Lines are also of the same importance in forming the gas flow in the outer layers of the expanding supernova envelope (Fig. 12). After the shock emerges on the presupernova surface at $t = 0.0823$ days, an expansion of the envelope begins, and it rapidly (during the first few days) passes to free expansion. In model N, the resonant scattering of radiation in numerous lines accelerates the outer layers up to velocities of $\approx 36\,000$ km s$^{-1}$, while in model Nlo, which disregards the line opacity, the velocity of the outer layers does not reach even $30\,000$ km s$^{-1}$. Additional line-induced acceleration takes place in the outermost layers with a mass of $\approx 10^{-6} M_\odot$.

### Limb Darkening

The solution of the radiation hydrodynamics equations (1)–(5), which contain only the angular moments of the radiation intensity, yields the behavior of these moments in time and space, while the solution of the time-independent transfer equation (10) yields the limb-darkening law. Using the derived angular dependence of the intensity of the radiation emerging from the supernova envelope in calculating the bolometric luminosity with the delay effect from formulas (12) and (13) proves to be a very important factor, as shown by a comparison with the isotropic radiation in model Nld (Fig. 13). During the first ∼25 days, when a well-defined photosphere exists, particularly at the CW formation phase, the emitted radiation is nearly isotropic, and the bolometric light curves for models N and Nld almost coincide. As the envelope expands and becomes optically thin, the continuum formation region gradually becomes more extended, while the degree of anisotropy of the emergent radiation increases. The increase of the latter with time clearly shows the difference between the light curves for models N and Nld, which reaches particularly large values after the luminosity peak.

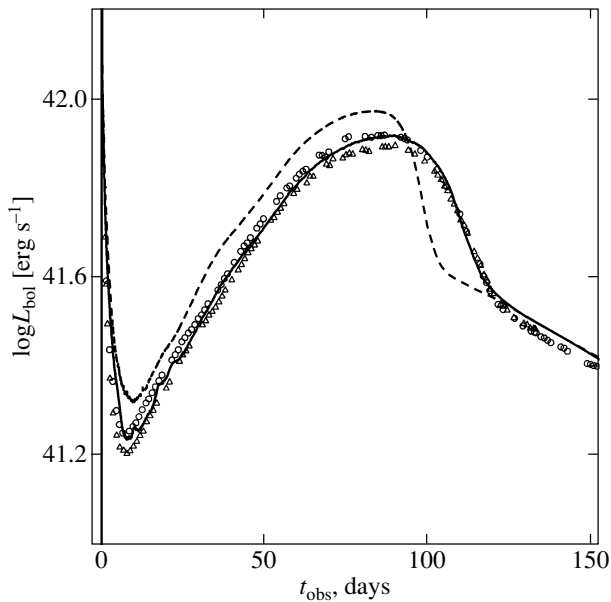

**Fig. 10.** Same as Fig. 4 for model N (solid line) and model Nlo—model N without line opacity (dashed line).

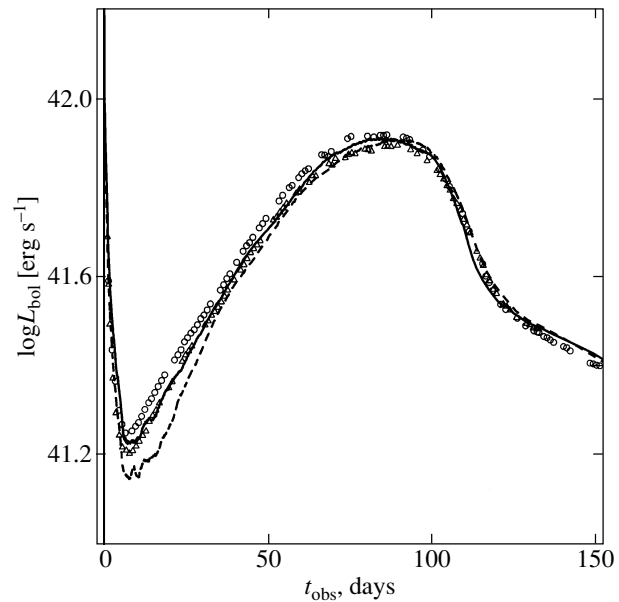

**Fig. 11.** Same as Fig. 4 for model Nce—model N with the LMC chemical composition and the explosion energy $E = 1.2 \times 10^{51}$ erg (solid line) and model Ncc—model N with the LMC chemical composition (dashed line).

Interestingly, the bolometric luminosity calculated by taking into account the limb-darkening law is higher and lower than the luminosity for isotropic radiation before and after the luminosity peak (Fig. 13), respectively. To explain this behavior of the bolometric luminosity, let us simplify formula (12) by assuming that the envelope is expanding freely, $R(t) \approx u_{\text{out}} t \approx u_{\text{out}} t_{\text{obs}} (1 + u_{\text{out}} \mu/c)$, and by fitting the emergent anisotropic radiation as $I(R(t), \mu) \approx I(\mu) \propto \mu^{\alpha}$, where $u_{\text{out}}$ is the velocity of the outer boundary of the supernova envelope, and $\alpha$ is a positive number that specifies the degree of anisotropy of the emergent radiation. Note the following two limiting cases: isotropic radiation ($\alpha = 0$) and a point radiation source ($\alpha \to \infty$). The above assumptions allow us to estimate the ratio of the bolometric luminosity $L(\alpha)$ calculated by taking into account the limb-darkening law described by the parameter $\alpha$ to the luminosity for isotropic radiation $L(0)$ at time $t_{\text{obs}}$:

$$\frac{L(\alpha)}{L(0)} \approx 1 + \frac{u_{\text{out}}}{c} \left( \frac{\alpha+2}{\alpha+3} - \frac{2}{3} \right) \frac{\partial \ln L(0)}{\partial \ln t_{\text{obs}}}. \quad (14)$$

As would be expected, this ratio depends on both the delay effect and the limb-darkening law. The correction on the right-hand side of formula (14) is on the order of $u_{\text{out}}/c$; it increases with increasing degree of anisotropy of the emergent radiation, while its sign is determined solely by the time derivative of the bolometric luminosity. The latter property entirely explains the observed behavior of the bolometric luminosity with an allowance made for the limb-darkening law before and after the luminosity peak.

### Influence of the Chemical Composition of the Outer Layers

The hydrodynamic models considered above have a chemical composition of the surface layers of the presupernova similar to that of the circumstellar matter. How do the basic parameters of the supernova explosion change if we substitute the chemical composition of the surface layers with the LMC chemical composition, which is characterized by a higher hydrogen abundance? Models Ncc and Nce, which have this property (see the table), give an answer to this question. Model Ncc shows that the passage to the LMC chemical composition in the surface layers accompanied by the corresponding increase in opacity causes the bolometric luminosity to decrease in the interval $t_{obs} \approx 5-80$ days (Fig. 11). In contrast, agreement with the observations of SN 1987A is achieved in model Nce, which has a higher explosion energy, $E = 1.2 \times 10^{51}$ erg, and a slightly modified chemical composition of the inner layers compared to model Ncc.

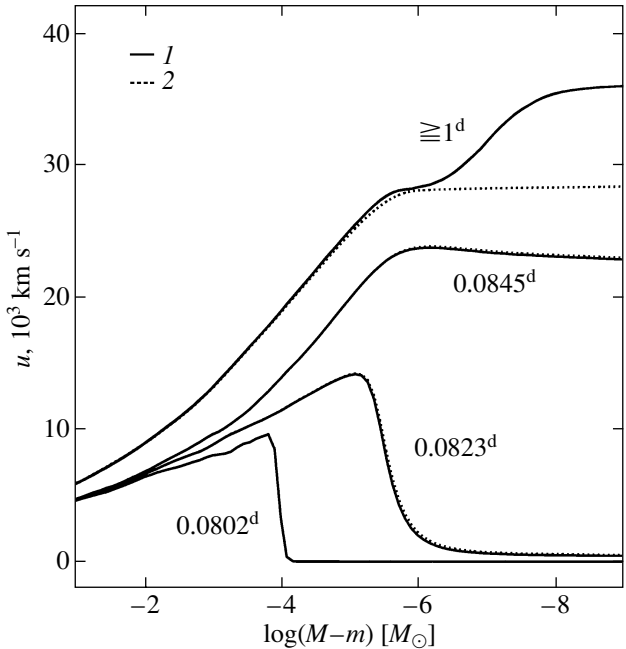

**Fig. 12.** Evolution of the velocity distribution in mass for model N (*1*) and model Nlo (*2*): at the time the shock approaches the presupernova surface $t = 0.0802$ days, at the time the shock emerges on the surface $t = 0.0823$ days, at $t = 0.0845$ days and, subsequently, at explosion times $t \geq 1$ days, when the envelope is already expanding freely. The mass is measured from the envelope surface.

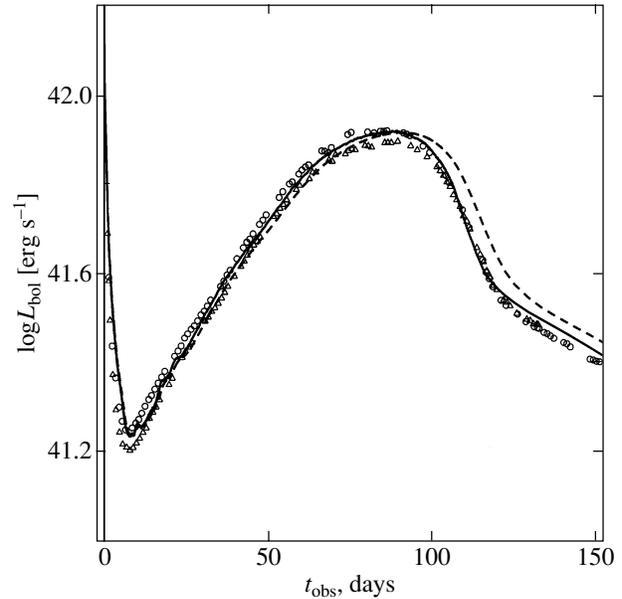

**Fig. 13.** Same as Fig. 4 for model N (solid line) and model Nld—model N without limb darkening (dashed line).

### DISCUSSION

Our study of the influence of the presupernova structure and the degree of $^{56}$Ni mixing on the bolometric light curve of SN 1987A in terms of radiation hydrodynamics in the one-group approximation by abandoning LTE and by taking into account nonthermal ionization and the contribution of lines to the opacity leads us to the following important conclusion: moderate (at velocities $u \leq 2500$ km s$^{-1}$) $^{56}$Ni mixing can explain the observed light curve if the density of the outer layers in the presupernova exceeds that in the evolutionary model of a single nonrotating star severalfold. Moderate $^{56}$Ni mixing is supported by the modeling of infrared emission lines at late stages, in the interval 200–700 days. Li *et al.* (1993) showed that the intensity of iron, cobalt, and nickel emission lines and their evolution with time are attributable to $^{56}$Ni mixing up to velocities of ~2500 km s$^{-1}$, while the same observations led Kozma and Fransson (1998) to conclude that the iron synthesized during the explosion was mixed to velocities of ~2000 km s$^{-1}$.

A similar serious constraint on the $^{56}$Ni mixing scales follows from an analysis of the so-called Bochum event in SN 1987A if the interpretation of it offered below is correct. This event consists in the formation of two peaks, blue and red, in the H$\alpha$ profile after ~20 days (Hanuschik and Dachs 1988). The blue peak can be explained by the nonmonotonic, spherically symmetric population distribution of the second hydrogen level in the atmosphere with the excitation minimum at velocities of ~4000–5000 km s$^{-1}$ (Chugai 1991), which is the result of nonstationary hydrogen recombination with a significant role played by the hydrogen neutralization processes involving H$^-$ and H$_2$ (Utrobin and Chugai 2002). The red peak is the result of a local enhancement of the hydrogen excitation from an asymmetric $^{56}$Ni ejection in the far hemisphere (Chugai 1991) with an absolute velocity of ~4600 km s$^{-1}$ (Utrobin *et al.* 1995). The spherically symmetric nonmonotonicity of the hydrogen excitation is very sensitive to the nonthermal excitation produced by $^{56}$Ni, and its existence for ~20–40 days implies that the $^{56}$Ni mixing at that time was within the photosphere and, accordingly, did not extend in velocity farther than $\sim$2500 km s$^{-1}$.

Thus, moderate $^{56}$Ni mixing in the envelope of SN 1987A receives confirmation both in spectroscopic observations at the photospheric phase and in nebular observations. The above analysis of the bolometric light curve convincingly shows that the outer layers of the presupernova have a density that is several times higher than that in the evolutionary model of a single nonrotating star. This important conclusion has the following direct implication: the presupernova structure required for the explosion of SN 1987A can be produced during the evolution of a single star in which, apart from the mass of the star on the main sequence and its chemical composition, stellar rotation may play a key role, or during the evolution of the presupernova in a close binary during the merger with the secondary component.

As must be the case, abandoning LTE and allowing for nonthermal ionization when solving the equation of state and determining the mean opacities and the thermal emission coefficient result in a significant difference between the gas temperature and the radiation temperature in the optically thin layers of the supernova envelope compared to the equilibrium case. This stresses the importance of the correct description of the energy exchange between matter and radiation when modeling supernova explosions.

Our study shows how important an allowance for the contribution of lines to the opacity is in the hydrodynamic modeling of SN 1987A. First, it has become clear that the contribution of lines to the opacity in an expanding envelope and the accurate description of the radiative transfer play a fundamental role in reproducing the observed shape of the bolometric light curve. Second, it has been established that disregarding the contribution of lines to the opacity introduces an error of $\sim$20% into the explosion energy, and that a similar error is possible when determining the mass of the ejected matter. Third, the resonant scattering of radiation in numerous lines accelerates the outer layers up to velocities of $\approx$36 000 km s$^{-1}$, while the velocities of these layers without an allowance for lines do not reach even 30 000 km s$^{-1}$; the additional line-induced acceleration affects the outer layers with a mass of $\approx 10^{-6} M_\odot$.

The next important conclusion is that proper calculations of the observed supernova luminosity require taking into account both the delay effects and the limb-darkening effects; the importance of the latter increases with time as the anisotropy of the emergent radiation increases. This conclusion is confirmed by our analytic estimate of the bolometric luminosity that includes the delay effects and the limb-darkening law. Accordingly, this estimate contains a explicit dependence on the time derivative of the bolometric luminosity and the degree of anisotropy of the emergent radiation.

In conclusion, let us note that substituting the chemical composition of the surface layers, which is similar to that of the circumstellar matter, with the LMC chemical composition, which is characterized by a higher hydrogen abundance, causes the bolometric luminosity to decrease in the interval $t_{\rm obs} \approx$ 5−80 days, and good agreement with the observations of SN 1987A is achieved with a higher explosion energy, $E = 1.2 \times 10^{51}$ erg.

## ACKNOWLEDGMENTS

I wish to thank S.I. Blinnikov and N.N. Chugai for their helpful discussions. I am grateful to A. Heger who provided the evolutionary model l20n2ae and comments to it. This work was supported in part by the Russian Foundation for Basic Research (project no. 01-02-16295).

*Translated by V. Astakhov*